\documentclass[twocolumn, preprintnumbers, endnote,nofootinbib,prl,aps,9pt,]{revtex4}
\usepackage[utf8]{inputenc}
\usepackage{epsfig,subfigure}

\usepackage{epstopdf}

\usepackage{graphicx}
\usepackage{amssymb}
\usepackage{amsxtra}
\usepackage{amsmath}
\usepackage{booktabs,multirow,tabularx}
\usepackage{slashed}
\usepackage{float}
\usepackage{placeins}
\usepackage{rotating}
\usepackage{lscape}
\usepackage{color}
\usepackage{hyperref}

\usepackage[Q=yes,pverb-linebreak=no]{examplep}

\DeclareGraphicsExtensions{.eps}
\graphicspath{{./}}

\newcommand{\lsim}{\lesssim}
\newcommand{\gsim}{\gtrsim}

\newcommand{\apr}{ A^\prime  }

\def\beq{\begin{equation}}
\def\bea{\begin{eqnarray}}
\def\eeq{\end{equation}}
\def\eea{\end{eqnarray}}
\def\beqnl{\begin{align}}
\def\endal{\end{align}}

\newcommand{\mone}{m_{\chi_{1}}}
\newcommand{\mtwo}{m_{\chi_{2}}}
\newcommand{\mdp}{m_{A^{\prime}}}
\newcommand{\aD}{\alpha_{D}}
\newcommand{\gminustwo}{g_{\mu} - 2}

\newcommand{\babar}{\texttt{BABAR}}
\newcommand{\slace}{\texttt{E137}}
\newcommand{\belle}{\texttt{BELLE II}}
\newcommand{\feynrules}{\texttt{Feynrules}}
\newcommand{\madgraph}{\texttt{MG5\_aMC@NLO}}
\newcommand{\lsnd}{\texttt{LSND}}
\newcommand{\miniboone}{\texttt{MiniBooNE}}
\newcommand{\na}{\texttt{NA64}}

\definecolor{red1}{cmyk}{0,1,1,0.1}
\definecolor{blue1}{cmyk}{1,0,0,0}

\DeclareFontFamily{U}{cbgreek}{}
\DeclareFontShape{U}{cbgreek}{m}{n}{
        <-6>    grmn0500
        <6-7>   grmn0600
        <7-8>   grmn0700
        <8-9>   grmn0800
        <9-10>  grmn0900
        <10-12> grmn1000
        <12-17> grmn1200
        <17->   grmn1728
      }{}
\DeclareFontShape{U}{cbgreek}{bx}{n}{
        <-6>    grxn0500
        <6-7>   grxn0600
        <7-8>   grxn0700
        <8-9>   grxn0800
        <9-10>  grxn0900
        <10-12> grxn1000
        <12-17> grxn1200
        <17->   grxn1728
      }{}

\makeatletter
\newcommand{\normalorbold}{%
  \ifnum\pdf@strcmp{\math@version}{bold}=\z@ bx\else m\fi
}
\makeatother

\begin{document}


\title{\boldmath Revisiting the Dark Photon Explanation of the Muon Anomalous Magnetic Moment}
\author{Gopolang Mohlabeng\footnote{email:gmohlabeng@bnl.gov}}

\affiliation{Physics Department, Brookhaven National Laboratory, Upton, New York 11973, USA}

\begin{abstract}
A massive $U(1)^{\prime}$ gauge boson known as a ``dark photon" or $A^{\prime}$, has long been proposed as a potential explanation for the discrepancy observed between the experimental measurement and theoretical determination of the anomalous magnetic moment of the muon ($g_{\mu} - 2$) anomaly. 
Recently, experimental results have excluded this possibility for a dark photon exhibiting exclusively visible or invisible decays. 
In this work, we revisit this idea and consider a model where $A^{\prime}$ couples inelastically to dark matter and an excited dark sector state, leading to a more exotic decay topology we refer to as a semi-visible decay.  
We show that for large mass splittings between the dark sector states this decay mode is enhanced, weakening the previous invisibly decaying dark photon bounds.
As a consequence, $A^{\prime}$ resolves the $g_{\mu} - 2$ anomaly in a region of parameter space the thermal dark matter component of the Universe is readily explained.
Interestingly, it is possible that the semi-visible events we discuss may have been vetoed by experiments searching for invisible dark photon decays. 
A re-analysis of the data and future searches may be crucial in uncovering this exotic decay mode or closing the window on the dark photon explanation of the $g_{\mu} - 2$ anomaly.
\end{abstract}

\maketitle

\section{Introduction\label{sec:intro}}
The anomalous magnetic moment of the muon $a_{\mu} \equiv (g_{\mu} - 2)/2$ remains to this day one of the few outstanding problems in particle physics.
A difference between theory and experiment of 
\begin{equation}
\Delta a_{\mu} \equiv a_{\mu}^{exp} - a_{\mu}^{th} = (274 \pm 73) \times 10^{-11} ,
\end{equation}
has resulted in a $\sim$ 3.7$\sigma$ discrepancy \cite{Bennett:2006fi, Patrignani:2016xqp} which is yet to be understood. While impressive agreement has existed between the Standard Model (SM) prediction and measurements on the electron's anomalous magnetic moment $a_{e}$ \cite{Aoyama:2012wj}, a recent improvement in the determination of the fine structure constant $\alpha$ from atomic Cesium measurements \cite{Parker:2018vye} has pushed the discrepancy in $\Delta a_{e}$ from $\sim 1.7 \sigma$ to $\sim 2.4 \sigma$ with opposite sign to that of the muon \cite{Hanneke:2008tm, Hanneke:2010au, Aoyama:2017uqe} \footnote{We note here that the contribution of the dark photon to lepton magnetic moments is positive \cite{Fayet:2007ua, Pospelov:2008zw}. Therefore it cannot simultaneously explain the negative $g_{e} - 2$ and positive $g_{\mu} - 2$ anomalies outlined above. For a possible model which allows an explanation for both anomalies we refer the reader to Ref.~ \cite{Davoudiasl:2018fbb}}. \\

Crucially, in the case of the muon, important progress from both experiment and theory lies in the near future. The upcoming Fermilab E989 \cite{Grange:2015fou} and J-PARC E34 \cite{Iinuma:2011zz} experiments will attempt to lower the uncertainty of the BNL E821 result by a factor of $ \sim$4. In parallel, progress on the SM theory side is expected to lower the corresponding theoretical uncertainties \cite{Blum:2018mom, Keshavarzi:2018mgv, Izubuchi:2018tdd}. Finally, it is now well known that leptonic moments can be exquisite probes of beyond the SM (BSM) physics \cite{Czarnecki:2001pv}. 
In this Letter, we revisit the ``dark photon'' - a light $U(1)$ vector boson explanation of the $g_\mu - 2$ anomaly \cite{Fayet:2007ua, Pospelov:2008zw} and outline a way to test it in the near future.
Dark photon phenomenology has been studied quite extensively in the literature and constraints from various experimental programs have been placed on its mass and coupling ($\epsilon$) to the SM. 
Searches for dark photons that can explain $g_\mu - 2$  have looked for resonant production and decay of these. Therefore, one must make an assumption on the decay modes; 
\begin{itemize}

\item {\it Visible decays:} There has been tremendous interest over the last decade, and two kinds of searches have been pursued. The first possibility to be studied was visible decays of the new $U(1)$ boson into SM leptons. This possibility has been decisively excluded in the recent past by a host of experiments \cite{Bross:1989mp, Bjorken:1988as, Riordan:1987aw, Lees:2014xha, Batley:2015lha, Merkel:2014avp}. The last window for a dark photon explanation was closed by the \texttt{NA48} experiment via rare $\eta\rightarrow \gamma  A'$ decays \cite{Batley:2015lha}.

\item {\it Invisible decays:} The dark photon could instead decay into invisible states, such as Dark Matter (DM). Such a possibility is particularly intriguing, as these models can readily explain the DM's observed relic abundance through freeze-out \cite{Fayet:2004bw, Izaguirre:2015yja}. Here, typical searches for the dark photon look for missing energy or missing mass. The last window for a dark photon explanation to $g_\mu - 2$ was closed in 2017 by the \texttt{NA64} \cite{Banerjee:2016tad}, and \texttt{BABAR} \cite{Lees:2017lec} experiments. 

\end{itemize}
\begin{figure}[t]
\includegraphics[scale=0.4]{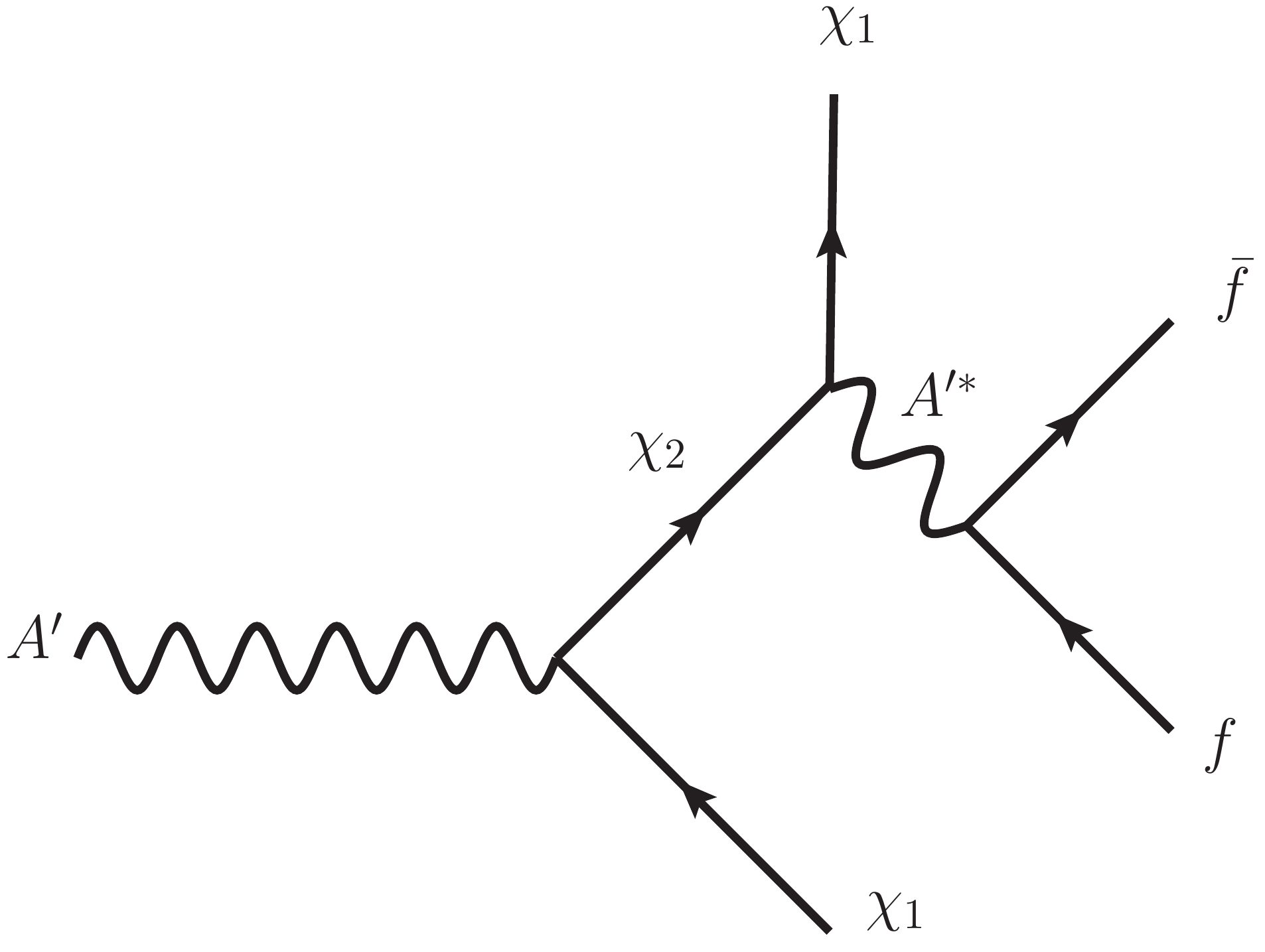}
\caption{Schematic diagram of dark photon decays in our scenario. Here the $\apr$ decays promptly 
to {\it both} visible and invisible particles, thereby evading the $\apr \to \bar f f$ resonance searches (where $f$ is a charged
SM fermion) and  missing energy searches for $\apr \to $ invisible decays. }
\label{cartoon}
\end{figure}
Here we examine a third possibility, namely the scenario where there exist dark sector states charged under some dark symmetry. If the symmetry is spontaneously broken, these states could have both dark symmetry preserving and breaking mass terms. 
In the limit that the symmetry breaking mass term is smaller than the preserving term, the dark sector states could exhibit off-diagonal interactions with the dark photon\footnote{This could be a realization of the inelastic dark matter scenario proposed in Ref.~ \cite{TuckerSmith:2001hy}.}.
 The heavier eigenstate can decay into the lighter state in conjunction with some visible SM states. 
In Fig.~\ref{cartoon} we illustrate how the dark photon may decay if it couples inelastically to two dark sector states $\chi_{1}$ and $\chi_{2}$, where the $\chi_{1}$ state is stable and forms the thermal DM component in the Universe. We refer to the decay in Fig.~\ref{cartoon} as the semi-visible decay mode of the dark photon\footnote{For various extentions of this model, see Refs.~\cite{Morrissey:2014yma, Darme:2017glc, Darme:2018jmx}}. 

\section{Model Setup \label{sec:model}}
In this work we focus on a well motivated massive gauge boson ($A^{\prime}$) charged under a hidden $U(1)^{\prime}$ extension of the SM. This hidden sector interacts with the SM through a gauge invariant and renormalizable $\rm photon-A^{\prime}$ kinetic mixing term, $\mathcal{L}_{ \rm mix} = \epsilon F^{\prime \mu \nu} F_{\mu \nu}$, where $\epsilon$ is the kinetic mixing parameter, and $F_{\mu\nu}$ and $F'_{\mu\nu}$ are the field strengths for the SM photon and dark sector photon (``dark photon") respectively.
After diagonalizing the kinetic mixing term, the interaction between the hidden and visible sectors is given by
\bea
\mathcal{L}_{\rm int} \supset  A^{\prime}_{\mu} (\epsilon e \mathcal{J}^{\mu}_{EM} + g_{D} \mathcal{J}^{\mu}_{D}),
\eea
where $\mathcal{J}^{\mu}_{EM}$ is the SM electromagnetic current and $\mathcal{J}^{\mu}_{D}$ is the dark sector current, coupling to $A^{\prime}$ with strength $g_{D} \equiv \sqrt{4\pi \alpha_{D}}$. 
In our representative model, $\mathcal{J}^{\mu}_{D}$ contains four-component fermionic states $\Psi = \begin{pmatrix} \eta & \xi^\dagger \end{pmatrix}$ ($\eta$ and $\xi$ are two-component Weyl fermions) which in general can couple to a dark scalar $h_{D}$ through a gauge-invariant Yukawa coupling of the form $y_D h_D \bar \Psi^c \Psi$. Dark symmetry breaking gives rise to Majorana mass terms $\frac{m_{\xi}}{2} \xi\xi$ and $\frac{m_\eta}{2} \eta \eta$ in addition to a symmetry-preserving Dirac mass term $m_D\eta \xi$.
Diagonalizing the Dirac and Majorana masses gives rise to mass eigenstates $\chi_1$ and $\chi_2$ (which are linear combinations of $\eta$ and $\xi$) and we obtain  
\beq
\mathcal{L} \supset \sum_{i = 1, 2} \bar \chi_i (i \slashed{\partial} - m_{\chi_{i}}) \chi_i - (g_D \bar\chi_2 \slashed{A'} \chi_1 + \rm{h.c.}),
\eeq
The general case $m_{\eta} \neq m_{\xi}$ results in both inelastic and elastic terms in the vector current, whereby the dark matter state $\chi_{1}$ and the ``excited'' state $\chi_2$, with mass splitting $\Delta \equiv \mtwo - \mone$, couple dominantly off-diagonally to the dark photon. However there is also a subdominant diagonal coupling, $\sum_{i = 1, 2} \bar \chi_i \slashed{A'} \chi_i$. 
In this scenario the thermal DM abundance is set by the annihilation of $\chi_1$ as well as co-annihilation of $\chi_1$ and $\chi_2$ into the SM. 
The former (Dirac fermion DM) is robustly excluded by CMB data for $m_{\rm DM} < 10$ GeV~\cite{Ade:2015xua}. 
Therefore in this work we focus primarily on the scenario in which $m_{\eta} = m_{\xi}$ such that there is only an off-diagonal inelastic coupling between $\chi_1$, $\chi_2$ and the dark photon (Pseudo-Dirac DM). 
Here the relic abundance is set by the co-annihilation of $\chi_1$ and $\chi_2$ as well as down-scattering and decay of $\chi_2$ into $\chi_1$ and SM states, which are safe from CMB bounds (see Refs.~\cite{Izaguirre:2015zva, Izaguirre:2017bqb} for more detailed information).

\section{Constraints \label{sec:const}}
The primary goal of this paper is to illustrate that in an iDM model with large mass splittings ($\Delta ~\gsim ~40\%$) we can significantly weaken the existing limits such that the previously excluded 2$\sigma$ dark photon explanation of the $\gminustwo$ anomaly is still viable, and in a region of parameter space the thermal relic dark matter abundance is readily explained.
To show this, we recast the invisibly decaying dark photon limits based on the following possibilities: 
\begin{enumerate}
\item In a beam dump, $\chi_{2}$ after it's production from prompt $A^{\prime}$ decay, is long-lived and can travel a distance to the detector, decaying into $\chi_{1} f \bar{f}$ inside the detector.
Alternatively, in a collider environment the dark photon may be produced through the reaction $e^{+} e^{-} \rightarrow  \gamma A^{\prime} $, with a subsequent prompt decay $A^{\prime} \rightarrow \chi_{1} \chi_{2}$. 
If $\chi_{2}$ is long lived it will decay outside the detector, resembling a monophoton signature.
On the other hand, $\chi_{2}$ may be short lived enough to decay inside the detector.
If the SM final states fall below the detector thresholds then the signal resembles a monophoton + $\slashed{E_{T}}$ signature. However if the SM decay products are above the detector thresholds then such an event topology may give rise to displaced tracks as well as missing momentum in the final state.
The probability for $\chi_{2}$ to decay either inside or outside the detector is determined by its decay width which may be approximated as
\begin{equation}
\Gamma(\chi_{2} \rightarrow \chi_{1} \bar{f} f) \sim \frac{4 \epsilon^{2} \alpha \alpha_{D} \Delta^{5}}{15 \pi m_{A^{\prime}}^{4}}.
\end{equation}
Larger values of $\Delta$ imply a higher probability of decaying inside the detector, thus evading elastic DM bounds.
\item Either states $\chi_{1,2}$ may up or down scatter in the detector material, producing a recoiling target signal.
\end{enumerate}
To obtain the results in our analysis, we defined a representative dark photon model coupled to fermionic iDM using the \feynrules ~package
\cite{Alloul:2013bka} and carried out all model simulations using $\madgraph$ \cite{Alwall:2011uj}. In what follows we will discuss the existing constraints on the model space.\\
\begin{figure*}[t]
\centering
\includegraphics[scale=0.48]{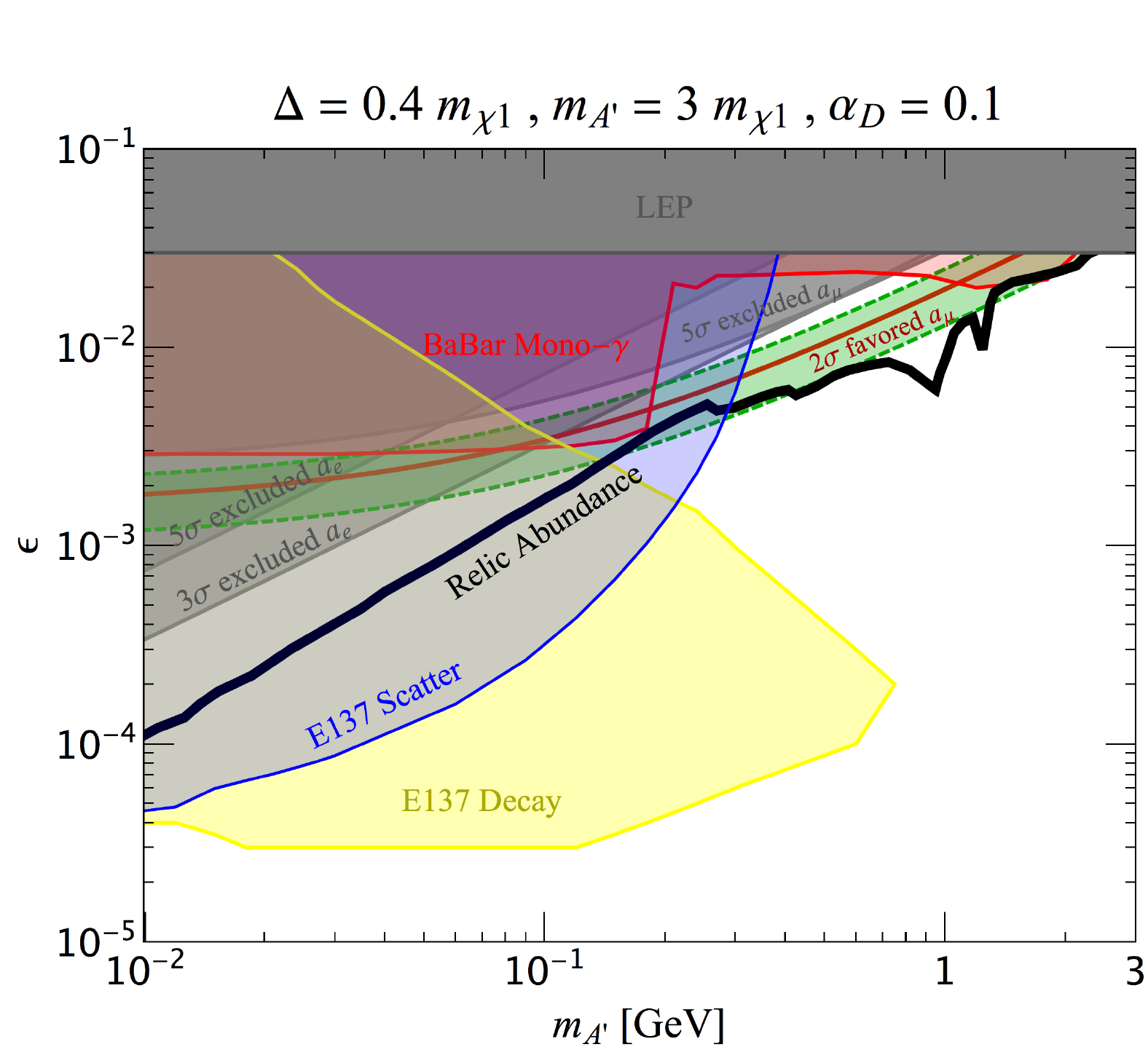}
\hspace*{0.3cm}
\includegraphics[scale=0.48]{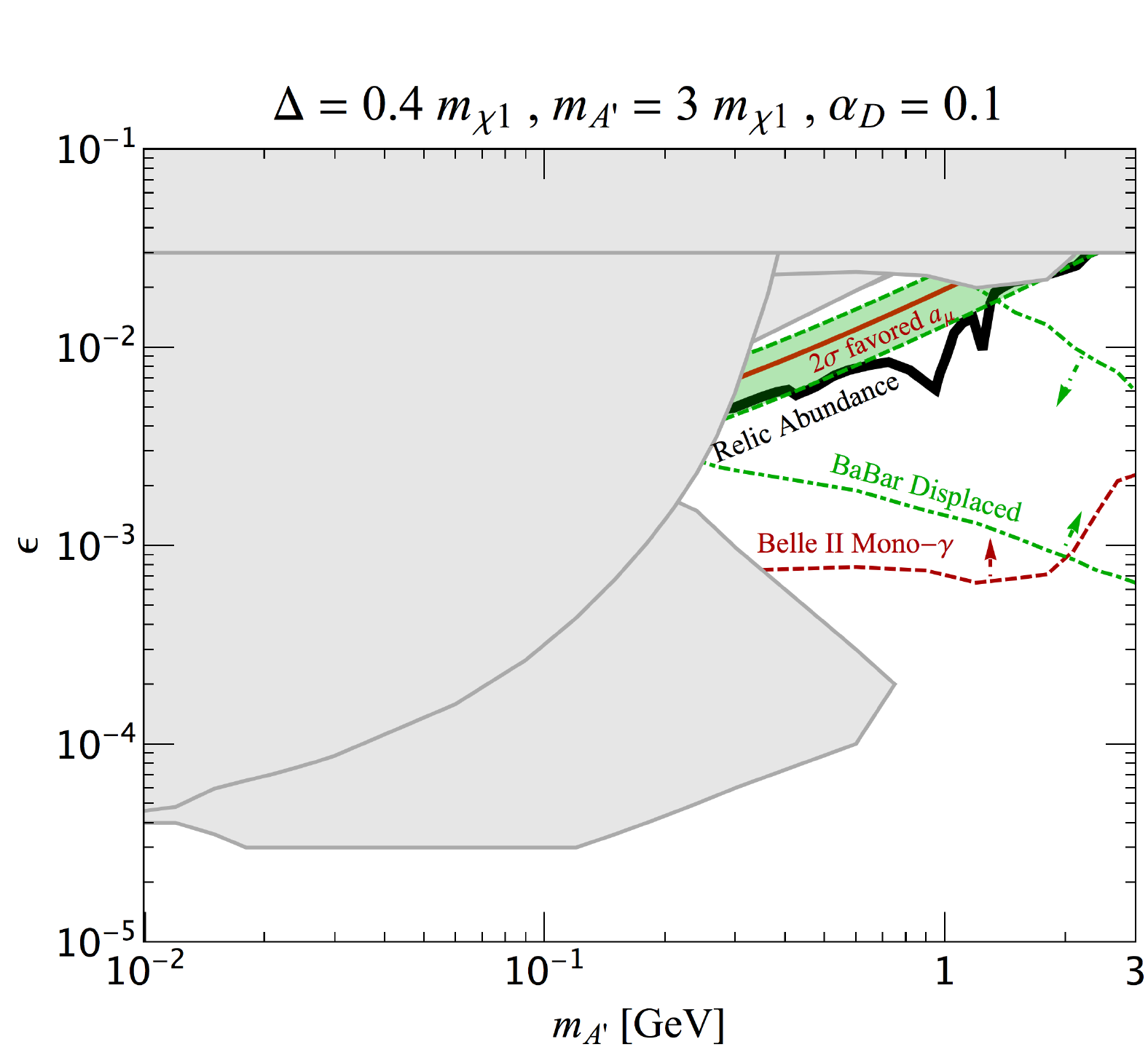}
\caption{Current limits on the dark photon semi-visible decay parameter space with benchmark parameters $\Delta = 0.4 ~\mone$, $\mdp = 3 ~\mone$ and $\alpha_{D} = 0.1$. On the left panel we place constraints on the $\epsilon$ vs $\mdp$ parameter space. On the right panel we include projections for a \belle~ monophoton search as well as a \babar~ displaced track re-analysis. We also color the various experimental constraints in grey for clarity and to bring attention to our region of parameter space. See text for details on the various bounds and projections.}
\label{fig:epsilon_alpha}
\end{figure*}

\textbf{B Factories}: We focus on the \babar~ monophoton + $\slashed{E_{T}}$ limits which have excluded an invisibly decaying dark photon for explaining $g_{\mu} - 2$.
In recasting this limit, we follow the analyses in Refs.~\cite{Essig:2013vha, Aubert:2008as,Izaguirre:2013uxa,Izaguirre:2015zva}. In our simulation, we reject events that have transverse impact parameter $b \lsim 350$ cm and have final state leptons with $p \gsim 60$ MeV. In addition, we adopt $\mathcal{L} = 53 fb^{-1}$ as used in the recent experimental search \cite{Lees:2017lec}.
For the monophoton + $\slashed{E_{T}}$ + displaced lepton projection we require a monophoton trigger as well as a selection on two displaced leptons with momentum $p \gsim 60 ~\rm MeV$. We also select events with transverse impact parameter between 1 cm and 50 cm and require a lepton reconstruction efficiency of $50\%$ (see Refs.~\cite{Izaguirre:2015zva, Aubert:2001tu, Allmendinger:2012ch} for more details, whose prescription we have followed here). 
For future reach, we consider the \belle~ experiment, which is projected to have a total integrated luminosity of 50 $ab^{-1}$ \cite{Kou:2018nap}. We assume a dedicated \belle ~ monophoton trigger and estimate its sensitivity by appropriately rescaling the \babar~ result by 50 $ab^{-1}$. 

\textbf{Fixed Target experiments}: \\ \textit{Electron beam dumps} - We study the \slace~ experiment at SLAC \cite{Bjorken:1988as} which offers sensitivity to the iDM signature. Following the discussion in Refs.~\cite{Batell:2014mga, Izaguirre:2017bqb, Izaguirre:2015yja}, we recast the \slace~ limit based on a non-zero probability that $\chi_{1,2}$ travelled to the detector and scattered off electron targets in the detector.
Alternatively, $\chi_2$ has some probability of surviving to the detector and decaying inside. For both signatures we require that the final state electrons have energy $ E_{e} \gsim 1$ GeV and fall within the detector's geometric acceptance. \\
\textit{Proton beam dumps} - These offer complementary probes of the iDM model. One such example is the \lsnd~  experiment \cite{deNiverville:2011it}. 
The relatively lower beam energy (800 MeV) means the dark photon is produced dominantly from the reaction $ \pi^{0} \rightarrow \gamma A^{\prime}$.
This implies that the \lsnd~ constraints for both scattering and decay signatures are kinematically limited by $\mone + \mtwo \equiv 2 \mone + \Delta~ \textless ~m_{\pi^{0}}$. For the purposes of this work, we focus on $m_{A^{\prime}} ~\textgreater ~m_{\pi^{0}}$ and do not discuss these bounds further.\\ 
On the other hand, the \miniboone~ experiment at Fermilab with a higher beam energy of 8 GeV is potentially sensitive to our model. 
Interestingly, the elastic dark matter scenario has been excluded in the $\gminustwo$ region, for a choice of $\alpha_{D}$ \cite{Aguilar-Arevalo:2017mqx, Aguilar-Arevalo:2018wea}. However the inelastic dark matter reach for both scattering and decay signatures, especially in the large mass splitting limit, is yet to be determined.
We leave this interesting possibility for a future dedicated study.\\
\textit{Missing momentum experiments} - These provide another potential avenue to search for the iDM signature. In particular, the \na~ experiment at CERN which triggers on missing momentum events and can search for the monophoton signature produced in the iDM model. The elastic dark matter limit, giving rise to the invisible decay signature is partially excluded by \na~\cite{Banerjee:2016tad} in the $\gminustwo$ region. However, the sensitivity of \na~ to the iDM model has not been studied and is left for a future work. \\

\section{Results \label{sec:res}}

In Fig.~\ref{fig:epsilon_alpha}, we plot the kinetic mixing parameter $\epsilon$ vs the dark photon mass $\mdp$, for benchmark values of $\Delta = 0.4 ~\mone$, $\mdp = 3 ~\mone$ and $\aD = 0.1$.  
On the left panel, the top grey shaded region represents the model independent upper bound extracted from electroweak precision observables at LEP and LHC \cite{Hook:2010tw, Curtin:2014cca}. The lighter grey shaded area is the region excluded by the muon g-2 experiment at 5$\sigma$ confidence level.
The green shaded band bounded by the green dashed lines is the 2$\sigma$ allowed region for the muon g-2 anomaly with the red solid line representing the central value \cite{Bennett:2006fi, Pospelov:2008zw}. 
We also include limits from the anomalous magnetic moment of the electron taking into account the most recent determination of the fine structure constant $\alpha$ \cite{Parker:2018vye}. 
We use the value of 
\beq
\Delta a_{e} = (-87 \pm 36.3) \times 10^{-14},
\label{eq:ae}
\eeq
derived in Ref.~\cite{Davoudiasl:2018fbb}. 
To set the bound from $g_{e} - 2$, we require that the $A^{\prime}$ contribution be no larger than $5\sigma$ from the central value (Eq.~\ref{eq:ae}), in the positive direction, translating to $a_{e}~ \textless~1.09 \times 10^{-12} $. For comparison we also include a more conservative bound of $3\sigma$ giving rise to $a_{e}~ \textless~2.2 \times 10^{-13} $. Interestingly the conservative $3\sigma$ $a_{e}$ bound covers the $5\sigma$ excluded $g_{\mu} - 2$ region, but not the $2\sigma$ allowed region of interest in this work.
The blue shaded area takes into account the bound from the SLAC \slace ~experiment \cite{Bjorken:1988as} assuming up/down-scattering of $\chi_{1,2}$ with SM particles in the detector. The yellow shaded bound is a recast of the \slace ~result accounting for long-lived $\chi_2$ particles traveling to and decaying inside the detector. 
The black line represents the thermal relic abundance of $\chi_{1}$ requiring $\Omega_{\chi_{1}} h^{2} \sim 0.12$ \cite{Aghanim:2018eyx}. For larger values of $\mone$ we take into account coannihilation to SM hadronic final states, hence the spikes in the higher mass region (please see Refs.\cite{Izaguirre:2015zva, Izaguirre:2017bqb,Ilten:2018crw} for more detailed information).
\begin{figure*}[t]
\centering
\includegraphics[scale=0.48]{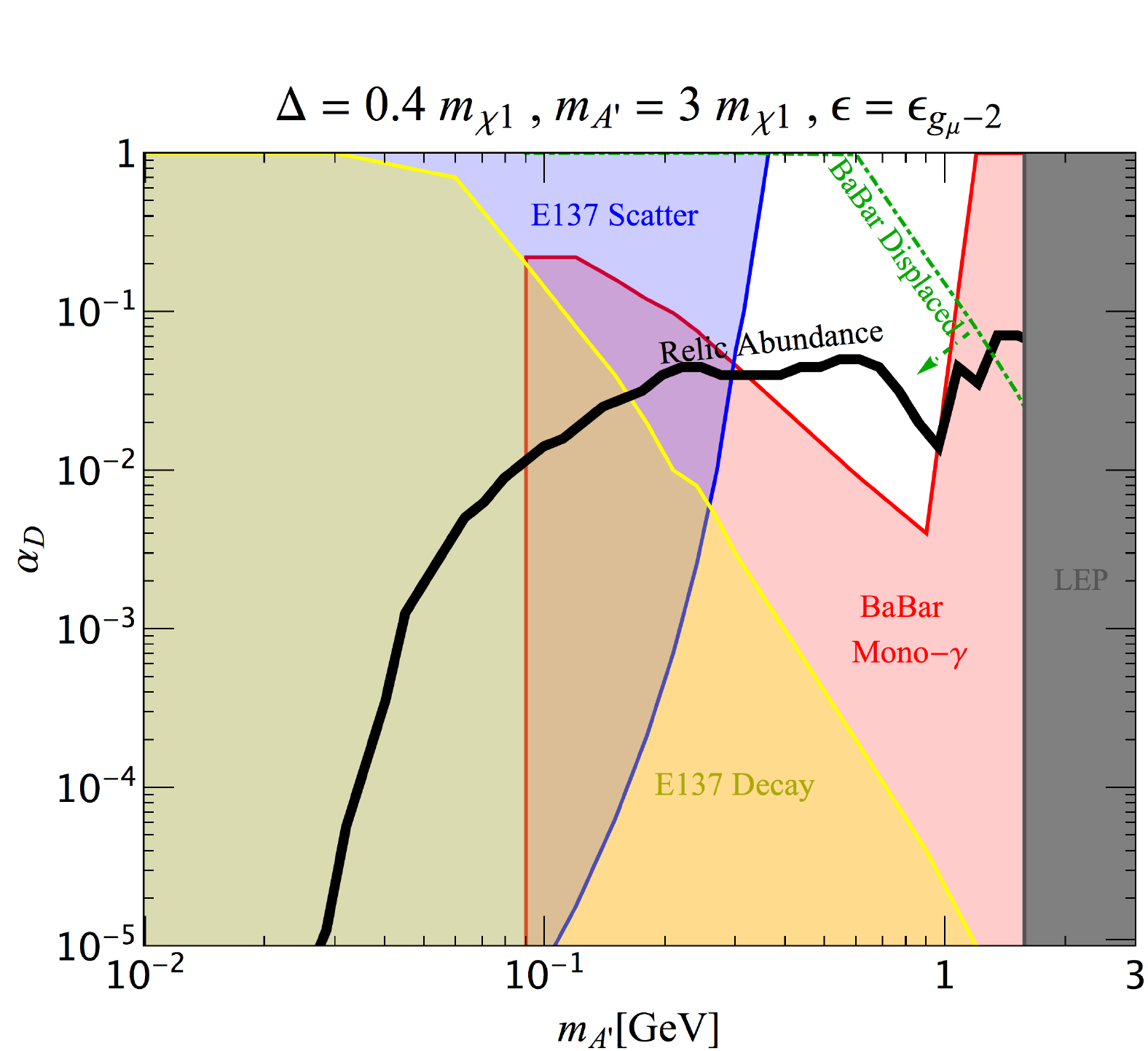}
\hspace*{0.3cm}
\includegraphics[scale=0.47]{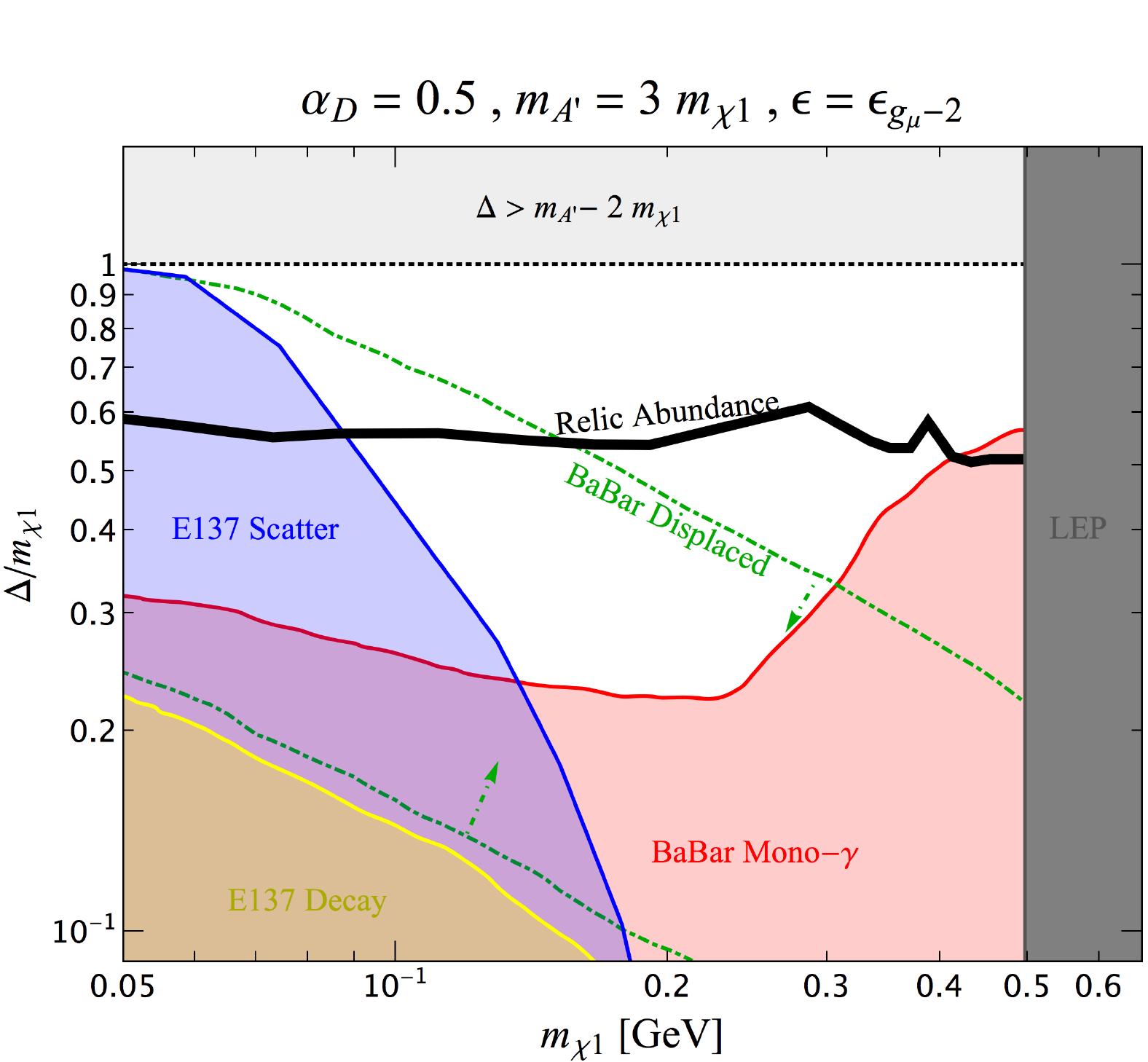}
\caption{Parameter space compatible with a dark photon decaying into thermal inelastic DM as an explanation for the Muon g-2 anomaly. On the left panel we plot $\alpha_{D}$ vs $m_{A^{\prime}}$ for a 40$\%$ mass splitting and $\mdp = 3 ~\mone$. On the right panel we plot the ratio of mass splitting to $\mone$ as a function of $\mone$ for benchmark parameters $\alpha_D = 0.5$ and $\mdp = 3 ~\mone$. See text for details.}
\label{fig:delta}
\end{figure*}
\\
The red shaded region represents the \babar  ~monophoton bound on our parameter space. In this region $\chi_{2}$ is long-lived and decays outside the detector or maybe short-lived, but its decay products are below the \babar ~thresholds, resembling a monophoton and missing energy signature. 
The decay width of $\chi_{2}$ scales with $\Delta$ as $\Gamma_{\chi_{2}} \sim \Delta^{5}$. For our parameter choices the region $\mdp~ \gsim ~100$ MeV corresponds to large values of $\Delta$, increasing the probability that $\chi_{2}$ will decay inside the detector. Hence the \babar~ monophoton limit is weakened in this region, opening up the $2\sigma$ favored explanation for $\gminustwo$.
For clarity of our results, we represent all the bounds from the left panel, as the solid grey region in the right panel and include projections in our region of interest. 
A particularly striking indication of the semi-visible decay mode would be a monophoton + displaced track + missing energy signature which could be uncovered by a future re-analysis of the \babar ~data.
This is illustrated as the region bounded by the green dot-dashed lines and marked by the green arrows. Interestingly, both the $\gminustwo$ and thermal relic lines fall within the region projected to be uncovered by the displaced track search.
Finally, the darker red dashed bound is the projection for the \belle ~experiment, the arrow indicates the region of parameter space \belle~ is expected to cover.

In Fig.~\ref{fig:delta} we set $\epsilon$ as the central value required to explain the $\gminustwo$ anomaly (i.e. the central red line in Fig.~\ref{fig:epsilon_alpha}). On the left panel we plot the dark sector coupling $\alpha_{D}$ as a function of $\mdp$, with $\mdp = 3 ~\mone$ and $\Delta = 0.4~\mone$. 
The white region indicates the parameter space available, while the color shaded regions are excluded by the various experimental contraints as in Fig~\ref{fig:epsilon_alpha}. 
For the parameter choices in this discussion, it is interesting to see that the iDM model can simultaneously explain the DM thermal relic abundance and $\gminustwo$ anomaly for $300~ \rm MeV \lsim \mdp \lsim 1$ GeV.
A \babar~ displaced re-analysis would uncover most of this parameter space, up to large values of $\alpha_D$ where we start reaching perturbativity limits.\\
On the right panel of Fig.~\ref{fig:delta} we plot the ratio of mass splitting $\Delta$ to the DM mass $\mone$ as a function of $\mone$, with $\alpha_{D} = 0.5$.
Here the lighter grey region bounded by the horizontal dotted line at $\Delta/\mone \sim 1$ is a kinematic limit in which $\mone + \mtwo~ \textgreater~ \mdp$ i.e. $A^{\prime}$ is produced off-shell.
The unshaded area corresponds to the parameter space available for explaining the $\gminustwo$ anomaly. Also shown is the relic abundance line corresponding to a thermal relic $\chi_{1}$. The projected sensitivity of \belle~ in the parameter regions of Fig.~\ref{fig:delta} (with our choices of parameters) is nearly the same as the \babar~ region and thus we do not include it. We see that for larger mass splittings and choices of $\alpha_D$, we are able to explain DM and $\gminustwo$ simultaneously. We also show that part of the available thermal relic space would be uncovered with a displaced track search at \babar.

\section{Discussion and conclusions}
In this work, we have shown that a dark photon coupled to inelastic dark matter can explain both the $\sim 3.7 ~\sigma$ discrepancy observed in the anomalous magnetic moment of the muon, and the thermal dark matter of the Universe simultaneously. 
To illustrate this, we recast previous bounds on an invisibly decaying $A^{\prime}$ and found that in the presence of large mass splittings, the \babar~ monophoton limit is weakened, especially in the region the 2$\sigma$ $A^{\prime}$ explanation for the $g_{\mu}-2$ anomaly is allowed. 
Interestingly, the semi-visible decay channel we discuss here could be uncovered by a re-analysis of the current \babar~ data, searching for a monophoton + displaced track + $\slashed{E_{T}}$ signature, though admittedly it is not clear how complex this re-analysis may be. 
Furthermore, it would be interesting to investigate whether a recast of the \na~\cite{Banerjee:2016tad, Izaguirre:2014bca}, \miniboone~\cite{Aguilar-Arevalo:2017mqx, Aguilar-Arevalo:2018wea, deNiverville:2016rqh} and \texttt{NO$\nu$A} \cite{deNiverville:2018dbu} studies for invisible $A^{\prime}$ decays would
close off the allowed $\gminustwo$ window or leave it open; this possibility is left for future work. 
Finally, we note that future experiments such as \texttt{LDMX}\cite{Akesson:2018vlm}, \texttt{BDX}\cite{Battaglieri:2016ggd} (which have the potential to improve upon both E137 scattering and decay searches) as well as \texttt{$JSNS^{2}$}, \texttt{SeaQuest} \cite{Berlin:2018bsc, Jordan:2018gcd} and possibly \texttt{FASER} \cite{Berlin:2018jbm} could be sensitive to this parameter region and either confirm or completely close the window on the dark photon explanation of the $\gminustwo$ anomaly.

\bigskip

\textbf{Acknowledgements.}
GM would like to express special gratitude to both Eder Izaguirre and Gordan Krnjaic for their earlier collaboration and all their guidance and  encouragement throughout this work. He would also like to thank Hooman Davoudiasl, Rouven Essig, Sam Homiller, Bill Marciano, Aaron Meyer and David Morrissey for very helpful discussions. GM is supported by the United States Department of Energy under Grant Contract DE-SC0012704.

\bibliography{IDM_DP}

\end{document}